\documentclass[twocolumn,showpacs,amsfonts,aps,prc,nofootinbib,floatfix,%
superscriptaddress]{revtex4}

\usepackage{float}

\usepackage{amsmath}
\usepackage{bm}
\usepackage{graphicx}
\usepackage{xcolor}

\usepackage{epsfig}
\newcommand{\beq}{\begin{equation}}
\newcommand{\eeq}{\end{equation}}
\newcommand{\bea}{\vspace{0.25cm}\begin{eqnarray}}
\newcommand{\eea}{\end{eqnarray}}

\newcommand{\ro}{\mbox{{\boldmath
$\rho$}}}

\newcommand{\bb}{{{\bf b}}}

\def\lsim{\mathrel{\rlap{\lower4pt\hbox{\hskip1pt$\sim$}}
    \raise1pt\hbox{$<$}}}         
\def\gsim{\mathrel{\rlap{\lower4pt\hbox{\hskip1pt$\sim$}}
    \raise1pt\hbox{$>$}}}         

\begin{document}


\title{\Large\bf
  Predictions for $R_{AA}$ in 5.36 TeV C+C, O+O, and Ne+Ne collisions
  at the LHC
}

\date{\today}

\author{B.G. Zakharov}

\address{
L.D.~Landau Institute for Theoretical Physics,
        GSP-1, 117940,\\ Kosygina Str. 2, 117334 Moscow, Russia
}

\begin{abstract}
Experiments on collisions of light nuclei at $\sqrt{s}=5.36$ TeV 
have recently begun at the LHC.
In this regard we make predictions for nuclear modification factor $R_{AA}$
  in $5.36$ TeV C+C, O+O, and Ne+Ne collisions
for scenarios with and without quark-gluon plasma formation in $pp$ collisions.
We find a sizeable difference in $R_{AA}$ for these two scenarios,
which grows with decreasing atomic number.
This says that data on $R_{AA}$ for light nuclei
could potentially give information on the presence of jet quenching
in $pp$ collisions.
\end{abstract}
%

\maketitle
    {\bf I. Introduction.} The formation of a quark-gluon plasma (QGP)
    has been observed
in heavy ion collisions in experiments at RHIC and LHC.
One of the key signatures of the QGP formation in
$AA$ collisions is suppression of high-$p_{T}$ hadron spectra
(jet quenching) due to parton energy loss in the QGP
(for reviews, see  \cite{JQ_rev,QGP_sign}).
The dominant mechanism for parton energy loss in the QGP is the medium
induced gluon emission \cite{BDMPS1,LCPI1,W1,GLV1,AMY1}.

There are some indications that at the LHC energies a QGP 
can be formed in $pp/pA$ collisions as well. They include:
the observation of the ridge effect
in $pp$ collisions \cite{CMS_ridge,ATLAS_mbias}, 
the steep growth  of the strangeness  production in $pp$ collisions
at charged multiplicity density $dN_{ch}/d\eta\gsim 5$ \cite{ALICE_strange},
the anomalous multiplicity dependence of $\langle p_{T}\rangle$ \cite{Camp1}.
If a mini-QGP fireball is formed in $pp$ collisions, then
there must be some jet modification due to parton energy loss.
The QGP formation and jet quenching in small
systems have received
considerable experimental and theoretical attention in recent years
(see, e.g., a recent review \cite{UW_mQGP}). 

The medium suppression of high-$p_{T}$ hadrons in $AA$ collisions
is quantified by the nuclear modification factor $R_{AA}$, defined
as the ratio
\beq
R_{AA}=\frac{d^2N(AA\to hX)/dp_T^2dy}{N_{ev}
  \langle T_{AA}\rangle_{\Delta c} d^2\sigma(pp\to hX)/dp_T^2dy},
\label{eq:10}
\eeq
where $N(AA\to hX)$ is the hadron yield, $N_{ev}$ is the number of events, $\langle T_{AA}\rangle_{\Delta c}$ is the nuclear overlap function
for centrality bin $\Delta c$.
Centrality is experimentally determined
via charged hadron multiplicity. For heavy ion collisions, when the
multiplicity fluctuations are relatively small, centrality is tightly
correlated with the impact parameter, $b$ ($c(b)\approx\pi b^2/\sigma_{in}(AA)$
\cite{centrality}, except for very peripheral collisions).
In theoretical calculations of the $R_{AA}$ in the scenario without mini-QGP
formation in $pp$ collisions one can use for the reference $pp\to hX$
cross section in the denominator of (\ref{eq:10})
ordinary pQCD cross section $\sigma_{pQCD}(pp\to hX)/dp^2_{T}dy$.
For this theoretical nuclear modification factor 
we use the notation $R_{AA}^{st}$.
In the scenario with mini-QGP formation in $pp$
collisions, the real inclusive $pp$ cross section
differs from the pQCD one by the medium modification
factor $R_{pp}$, i.e.,
\beq
\frac{d^2\sigma_{}(pp\to hX)}{dp_{T}^2dy}=
R_{pp}\frac{d^2\sigma_{pQCD}(pp\to hX)}{dp_{T}^2dy}\,.
\label{eq:20}
\eeq
The $R_{pp}$ is an unobservable quantity. But
it modifies the theoretical $R_{AA}$, since in this scenario
the real $R_{AA}$ 
reads \cite{Z_pp_PRL,Z_pp13}
\beq
R_{AA}=\frac{R_{AA}^{st}}{R_{pp}}\,.
\label{eq:30}
\eeq

In \cite{Z_hl} we have performed a global analysis of the data
from RHIC and LHC on $R_{AA}$ in heavy ion collisions  
for scenarios with and without mini-QGP formation in $pp$ collisions.
We used the jet quenching scheme of \cite{RAA20} (adopted to use a temperature
dependent running $\alpha_s$ \cite{RAA20T} with a single free parameter)
based on the light-cone path integral (LCPI) approach to
the induced gluon emission \cite{LCPI1,LCPI2004,PT_JETP}.
The results of \cite{Z_hl} show that data on
$R_{AA}$ in heavy ion
collisions can be described equally well
in the scenarios with and without mini-QGP formation in $pp$ collisions.
But for 7 TeV O+O collisions it was found that  $R_{AA}$ 
in the scenario with mini-QGP formation in $pp$ collisions turns out to be
noticeably larger than that in the scenario without mini-QGP formation
in $pp$ collisions. 
This occurs due to the presence of the factor
$R_{pp}$ in the denominator of (\ref{eq:30}). 
The results of \cite{Z_hl} indicate that measuring $R_{AA}$ for light nuclei
is interesting not only from the point of view observation
of jet quenching in collisions of light nuclei,
but also for clarifying whether there is jet quenching in $pp$ collisions.

Recently experiments on light ion collisions have 
begun at the LHC for $\sqrt{s}=5.36$ TeV \cite{CERN-OO536}.
This makes it interesting to perform calculations of $R_{AA}$
in light ion collisions at this energy. 
In the present work we perform calculations of $R_{AA}$
in $5.36$ TeV C+C, O+O, and Ne+Ne collisions
that can be measured in the LHC experiments. 
We perform calculations of $R_{AA}$ for light nuclei
within the LCPI jet quenching scheme  of \cite{Z_hl}
for scenarios with and without mini-QGP formation in $pp$ collisions
(sometimes, for brevity, referred to as w/ mQGP and w/o mQGP scenarios). 
The strategy of the analysis is to fit
a single parameter in the parametrization of
$\alpha_s$ \cite{RAA20T} to heavy ion data,
and then to calculate $R_{AA}$
for light nuclei without free parameters.
The jet quenching scheme of \cite{Z_hl} has a number of features which are
very crucial for extrapolating the jet quenching predictions
from large systems to small ones:
(i) The LCPI approach to the induced  gluon emission
includes all rescatterings of fast partons
and treats accurately the finite-size effects. (ii)
The scheme accounts for the Coulomb effects in parton
rescatterings.
(iii) The radiative and collisional energy losses are calculated
for a realistic temperature dependent running $\alpha_s(Q,T)$.
Features listed in (i) are clearly necessary for a model to be applicable
for calculations of $R_{AA}$ both for large and small systems.
An accurate treatment of the Coulomb effects, pointed in (ii), is
important because for small systems the dominating
contribution to the induced gluon spectrum comes from
the $N=1$ rescattering term, which
 vanishes (for massless partons) in the oscillator approximation \cite{Z_OA}. 
Feature (iii) allows  to avoid the
ambiguities connected with the choice of $\alpha_s$ for small systems.
This is very important because the energy loss is
approximately $\propto \alpha_s^3$. 
Note that the above strategy has been implemented
for a successful description \cite{Z_Ipp,Z_IpA}
of the ALICE data \cite{ALICE_Ipp,ALICE_Ipp_PLB} on
the multiplicity dependence of the hadron-triggered medium modification
factors $I_{pp,pA}$ within the scheme of \cite{Z_hl}.
However, due to large experimental errors,
it is not yet possible to draw a definite
conclusion that the effect observed in \cite{ALICE_Ipp,ALICE_Ipp_PLB}
is a consequence of parton energy loss in a small QGP. 

{\bf II. The theoretical framework.}
In this section we summarize briefly the main features of the model.
A more detailed description can be found in \cite{RAA20,Z_hl}.

{\it A. Model of the QGP fireball.}
The model uses a uniform QGP density distribution in
the transverse plane.
We describe the QGP fireball evolution within Bjorken's 1+1D model
\cite{Bjorken}, which leads to proper time
dependence of the entropy density
$s(\tau)/s(\tau_0)=\tau_0/\tau$ with $\tau_0$ the thermalization
time. At $\tau<\tau_{0}$ we take $s(\tau)=s(\tau_0)\tau/\tau_0$.
As in \cite{Z_hl}, for a basic version we use $\tau_{0}=0.5$ fm.
To understand the uncertainties related to the choice of $\tau_0$, we
also perform calculations for  $\tau_0=0.8$ fm.

For $AA$ collisions the initial fireball entropy
density is defined through charged multiplicity density
$dN_{ch}(AA)/d\eta$ using Bjorken's relation \cite{Bjorken}
\beq
s_{0}=\frac{C}{\tau_{0} S_{f}}\frac{dN_{ch}(AA)}{d\eta}\,.
\label{eq:40}
\eeq
Here $S_{f}$ is the overlap area of the colliding nuclei,
and $C=dS/dy{\Big/}dN_{ch}(AA)/d\eta\approx 7.67$ \cite{BM-entropy}. 
We calculate $dN_{ch}(AA)/d\eta$ within the optical Glauber wounded
nucleon model with the binary collision term \cite{KN-Glauber}.
We use the Woods-Saxon nuclear density
$\rho_{A}(r)=\rho_{0}/[1+\exp((r-R_{A})/d)]$
with parameters $R_A$ and $d$ from \cite{WS25}:
($R_A,d)=(2.265,0.419)$, $(2.592,0.443)$,  and $(2.773,0.538)$ fm
for C, O, and Ne  nuclei,
respectively\footnote{We checked that for light nuclei the difference
between $R_{AA}$ for the harmonic oscillator
shell model nuclear density and for the Woods-Saxon one is very small.}.
As in our calculations
of the nuclear modification factor $I_{pA}$ \cite{Z_IpA},
we take into account that
two wounded nucleons participating in jet production  generate 
not the minimum bias $pp$ multiplicity density (as in the Glauber
model for minimum bias $AA$ collisions \cite{KN-Glauber,GL-MRSS}), but
the underlying event (UE) charged multiplicity density
$dN_{ch}^{ue}(pp)/d\eta$ in $pp$ jet events.
The latter is bigger than the minimum bias
multiplicity density by a factor of $\sim 2-2.5$ \cite{Field}.
We evaluate $dN_{ch}^{ue}(pp)/d\eta$ at $\sqrt{s}=5.36$ TeV
by interpolating the ATLAS 
data \cite{ATLAS_UE_Nch}: $dN_{ch}^{ue}(pp)/d\eta\approx 7.5$ and $13.9$
at $\sqrt{s}=0.9$ and $7$ TeV, respectively.
Assuming that $dN_{ch}^{ue}(pp)/d\eta\propto s^{\delta}$,
we obtain $dN_{ch}^{ue}(pp)/d\eta\approx 12.85$.
We use the fraction of the binary collisions $\alpha=0.13$,
supported by the Monte-Carlo Glauber calculations of
\cite{MCGL1,MCGL2,MCGL_Xe},
that describe very well data on the centrality dependence of the
midrapidity $dN_{ch}(AA)/d\eta$ for $0.2$ TeV Au+Au,
$2.76$ and $5.02$ TeV Pb+Pb, and $5.44$ TeV Xe+Xe collisions.
As in \cite{MCGL_Xe}, we determine the NSD minimum bias midrapidity
charged multiplicity density
and the NSD $pp$ inelastic cross section
by interpolating the ALICE results for $2.76$ and $7$ TeV $pp$ collisions
\cite{ALICE_nch541,ALICE4968}. This procedure gives at $\sqrt{s}=5.36$ TeV:
$dN_{ch}(pp)/d\eta\approx 5.397$ and 
$\sigma_{in}^{NSD}(pp)\approx 56.04$ mb.
With these parameters we obtain for the midrapidity multiplicity
density in minimum bias jet events:
$dN_{ch}(AA)/d\eta\approx 51.02$, 64.15, and 82.13 for C, O, Ne nuclei,
respectively.

As for heavy ion collisions \cite{Z_hl}, we define $S_f$ for
overlap of two circles with radius
$R=R_{A}+kd$. For our basic version we take $k=2$.
In \cite{Z_hl} it was found that
for heavy ions $R_{AA}$ weakly depends on $k$ at 
$k\sim 2-3$. However, for light ion
collisions the variation of $R_{AA}$ with $k$ may be more pronounced.
To understand possible theoretical uncertainties in $R_{AA}$
we also perform calculations for $k=3$.
For $k=2$  and $\tau_0=0.5$ fm, we obtain for the initial QGP temperature in minimum bias collisions
(evaluated via the average entropy density)
for the ideal gas (the lattice \cite{t-lat}) entropy:
$T_0=258(286)$, $263(291)$, and $261(289)$ MeV  for C, O, and Ne nuclei,
respectively.

To calculate the QGP entropy for $pp$ collisions
we again use Bjorken's \cite{Bjorken} relation between entropy
and multiplicity. For the charged multiplicity density we take 
the UE charged multiplicity density $dN_{ch}^{ue}(pp)/d\eta$.
We take $S_f=\pi R_f^2$, where $R_f$ is the effective radius of
the mini-QGP fireball for $pp$ collisions. The $R_f$
may be viewed as an average radius for the whole range of the impact
parameter. 
Using the multiplicity dependence of $R_f$ obtained within
the IP-Glasma model \cite{glasma_pp,RPP} (see \cite{Z_hl} for details)
we obtain $R_f=1.493$ fm for 5.36 TeV $pp$ collisions.
With this value of $R_f$, we obtain for the initial temperature of the
mini-QGP fireball at $\tau_0=0.5$ fm: 
$T_{0}\approx 227\,\,(256)$ MeV 
for the ideal gas (the lattice \cite{t-lat}) entropy.
The possible uncertainties in the predictions of the IP-Glasma model for $R_f$
cannot lead to sizeable uncertainties in $R_{pp}$, since its variation
with $R_f$ is very small. This occurs due to a compensation 
between enhancement of the energy loss caused by increase of the 
QGP size and its suppression caused by reduction of the QGP density.
Note that, for the same reason, $R_{AA}$ for light nuclei also turns out to
be practically insensitive to the possible uncertainties in parameters of the
Woods-Saxon nuclear densities.  

A remark on the event-by-event fluctuations of the QGP density,
that are ignored in our model. For heavy ion collisions
the event-by-event fluctuations are relatively small (except for very
peripheral collisions).
However, for small systems the event-by-event fluctuations
of the total QGP entropy and the geometrical fluctuations
in the transverse plane (lumpiness) may be large.
Nevertheless, one can expect that
the effect of the event-by-event fluctuations
on $R_{AA}$ (and $R_{pp}$) should be small. The point is that
for small systems the
induced gluon spectrum is dominated by the $N=1$ rescattering term,
which is a linear functional of the 
medium density profile along the fast parton trajectory. This
fact guarantees that the effect of the event-by-event fluctuations
should be small. Also, one should bear in mind that for light
ion collisions (and even more so for $pp$ collisions)
we have a situation when the typical formation length $L_f$ for induced
gluon emission is comparable to
the parton path length $L$ in the QGP
or larger\footnote{For the $N=1$ rescattering
  regime $L_f\sim 2\omega_g/m_g^2$ \cite{Z_pp13}. Then, taking $m_g\sim 400$ MeV
  \cite{LH}, for gluon energy
  $\omega_g\gsim 1-2$ GeV we obtain $L_f\gsim 2.5-5$ fm.}. In this regime
the induced gluon spectrum should be sensitive to the total amount
of the matter along the fast parton trajectory in the QGP fireball
and not to the local fluctuations of the matter density.

We conclude this section with a comment
on the transverse flow effects (neglected in our model).
In \cite{BMS_flow} it was shown within the BDMPS formalism \cite{BDMPS1},
that the transverse flow corrections to predictions obtained using the
Bjorken model should be small.
A detailed numerical analysis of the flow effects 
performed in \cite{BG_flow} confirmed that the effect of the QGP transverse
expansion on the medium suppression of high-$p_T$ hadrons is small.
From the results of \cite{BG_flow} for $R_{AA}$ in 0.2 TeV
Au+Au collisions for
$20-30$\% centrality bin
one can conclude that the flow correction to $R_{AA}$ for light nuclei
(and to $R_{pp}$) should be relatively small (see discussion at the end of
section 2 of \cite{Z_hl}).

{\it B. Jet quenching scheme.}
The standard theoretical nuclear modification factor $R_{AA}^{st}$ 
can be written as
\beq
R_{AA}^{st} = \frac{\left\langle
  d^2\sigma_m(NN\to hX)/dp^2_{T}dy\right\rangle_{AA}}
{   d^2\sigma_{pQCD}(pp\to hX)/dp^2_{T}dy}\,,
\label{eq:50}
\eeq
where $d^2\sigma_m(NN\to hX)/dp^2_{T}dy$ is
the medium modified inclusive $NN$ cross sections, 
$\langle \dots\rangle_{AA}$ denotes averaging over the impact parameter
vector $\bb$,
the transverse coordinate vector $\ro_j$ of the jet production point
and the jet azimuthal angle $\phi$.
The formula (\ref{eq:50}) defines the theoretical nuclear modification
factor corresponding to (\ref{eq:10}) in the w/o mQGP scenario.
To calculate the real $R_{AA}$
  in the w/ mQGP scenario one
  should divide (\ref{eq:50}) by the $R_{pp}$.

  The factor $R_{pp}$ can be written as  (similarly to (\ref{eq:50}))
\beq
R_{pp} =\frac{\left\langle d^2\sigma_m(pp\to hX)/dp^2_{T}dy\right\rangle_{pp}}{ 
  d^2\sigma_{pQCD}(pp\to hX)/dp^2_{T}dy}\,,
\label{eq:60}
\eeq
where $\langle \dots\rangle_{pp}$ denote averaging over the geometry of 
the jet production in $pp$ collisions.
For $pp$ collisions we perform averaging over the jet production transverse
coordinates and the jet azimuthal angle for the central $pp$ collisions.
This seems to be reasonable since the jet production
is dominated by the almost head-on collisions.
We use the distribution of hard processes in the impact parameter plane 
obtained with the quark distribution from the MIT bag model
(we assume that the hard gluons follow approximately the same 
distribution in the transverse plane).

Similarly to the ordinary pQCD formula for inclusive hadron cross section, 
we write the medium modified inclusive $NN$ cross section 
via the medium modified fragmentation functions (FFs) $D_{h/i}^{m}$ and the
pQCD inclusive parton cross section for $N+N\to i+X$ process as
\bea
\frac{d^2\sigma_m(NN\rightarrow hX)}{dp^2_{T} dy}=
\int_{0}^{1} \frac{dz}{z^{2}}
D_{h/i}^{m}(z, Q) \nonumber\\
\times \frac{d^2\sigma_{pQCD}(NN\rightarrow iX)}{dp_{T}^{i2} dy}
\,
\label{eq:70}
\eea
with $p_{T}^{i}=p_T/z$.
The FF $D_{h/i}^{m}$ is defined  
as a $z$-convolution \cite{RAA08}
\beq
D_{h/i}^{m}(Q)\approx D_{h/j}(Q_{0})
\otimes D_{j/k}^{in}\otimes D_{k/i}^{DGLAP}(Q)\,,
\label{eq:80}
\eeq
where $D_{k/i}^{DGLAP}$ is the DGLAP FF for $i\to k$ process,
$D_{j/k}^{in}$ is the in-medium $j\to k$ FF,
and  $D_{h^/j}$ is the FFs for $j\to h$ hadronization transition.
We calculate the DGLAP FFs
$D_{k/i}^{DGLAP}$ using the PYTHIA event generator \cite{PYTHIA}.
For FFs $D_{h/j}$ we use the KKP
\cite{KKP} parametrization  with virtuality $Q_0=2$ GeV.
For $AA$ collisions we account for the nuclear modification of the PDFs
with the LO EPS09 correction \cite{EPS09}.

In our jet quenching scheme the FFs $D_{j/k}^{in}$, are calculated 
in the approximation
of the independent gluon emission \cite{RAA_BDMS}
supplemented by the momentum and the flavor sum rules
(the details can be found in \cite{RAA08,RAA20}).
The one gluon spectrum used in calculating $D_{j/k}^{in}$
is calculated using the method of \cite{Z04_RAA},
which resums all rescatterings of the fast partons in the QGP
for a realistic interaction with the QGP constituents.
The effect of the collisional energy loss
(calculated within the approach of \cite{Z_coll})
is treated as a perturbation to the
radiative mechanism (see \cite{Z_hl} for details).  

The induced gluon spectrum and the
collisional energy loss are calculated with running $\alpha_s$
parametrized in the form (supported by the lattice results
\cite{Bazavov_al1}) 
\beq
\alpha_s(Q,T)\, =\, \begin{cases}
\dfrac{4\pi}{9\log(\frac{Q^2}{\Lambda_{QCD}^2})}  & \mbox{if } Q > Q_{fr}(T)\;,\\
\alpha_{s}^{fr}(T) & \mbox{if }  Q_{fr}(T)\ge Q \ge cQ_{fr}(T)\;, \\
\frac{Q\alpha_{s}^{fr}(T)}{cQ_{fr}(T)} & \mbox{if }  Q < cQ_{fr}(T)\;, \\
\end{cases}
\label{eq:90}
\eeq
with $Q_{fr}=\kappa T$, $c=0.8$, $Q_{fr}(T)=\Lambda_{QCD}\exp\left\lbrace
{2\pi}/{9\alpha_{s}^{fr}(T)}\right\rbrace$ (we take $\Lambda_{QCD}=200$ MeV).
The form (\ref{eq:90}) with $c>0$ gives $\alpha_s\propto Q$
at small $Q$. To understand the sensitivity of our
results to specific form of the $Q$-dependence of $\alpha_s$
in the soft region, we also perform calculations for $c=0$, i.e. for
a flat $\alpha_s$ at $Q<\kappa T$.
In both the cases we fitted $\kappa$ to
the LHC data \cite{ALICE_r502,ATLAS_r502,CMS_r502}
on $R_{AA}$ in 5.02 TeV Pb+Pb collisions.

Note that the successful description
of heavy ion data on
the nuclear modification
factors $R_{AA}$ \cite{Z_hl} and $I_{AA}$ \cite{Z_IAA} 
within our jet quenching model allows to conclude that the model
describes correctly the medium modified FFs $D_{h/i}^{m}(z, Q)$ in
the broad range of $z$ ($z\gsim (2-3)\text{GeV}/E$ with
$E$ the jet energy). However, $R_{AA}$ is most sensitive to the behavior of
$D_{h/i}^{m}(z, Q)$ at $z\gsim 0.5$.

{\bf III. Numerical results.}
In this section we present the numerical results for $R_{AA}$
in 5.36 TeV C+C, O+O, and Ne+Ne collisions for 0--10\%, 10--20\%,
0--30\%, and 0--100\% centrality classes.
We are fully aware of a limited applicability of the optical
Glauber model for description of the centrality dependence of $R_{AA}$
due to decorrelation of multiplicity and impact parameter
($c-b$ decorrelation). 
The problem with the $c-b$
decorrelation for light nuclei to a greater extent concerns the
experimental determination of $R_{AA}$
(due to the presence of the model dependent
factor $\langle T_{AA}\rangle_{\Delta c}$ in the denominator of (\ref{eq:10}))
than the uncertainties in theoretical calculations of $R_{AA}$.
Indeed, from the theoretical side, the $c-b$ decorrelation leads to fluctuations
of the multiplicity (i.e. the QGP density) for a given impact parameter.
But, as noted above, for light nuclei the dominant contribution
to parton energy loss comes from $N=1$ rescattering, for which
the effect of fluctuations should be small.
Measuring the minimum bias $R_{AA}$ 
(i.e. for $0-100$\% centrality interval) seems to be the most appropriate
for observation of the
jet quenching effects in collisions of light nuclei, because
$\langle T_{AA}\rangle_{0-100\%}$ equals
$A^2/\sigma_{in}^{AA}$ \cite{GL-MRSS}, that is insensitive to the $c-b$ decorrelation.

Besides the final state QGP effects, high-$p_T$ particle spectra in $AA$
collisions are also affected by the difference between the
free and bound nucleon PDFs. The uncertainties in the nuclear PDFs
translate to the uncertainties in the magnitude of the parton
energy loss in the QGP required for description of the experimental
$R_{AA}$. This problem may be especially serious for light nuclei
\cite{Z_hl,OOnpdf}, when the jet quenching effects
become small. To illustrate the effect of the nuclear
modification of the nucleon PDFs on the $R_{AA}$ we also present
the results for the no-quenching nuclear modification factor
$R_{AA}^{pdf}$ obtained ignoring the parton energy loss in the QGP.

The values $\tau_0=0.5$ fm and $k=2$ seem to be most reasonable
choice for description of the QGP fireball. 
However, to understand the stability of theoretical predictions
for $R_{AA}$ in light ion collisions to variation of $\tau_0$ and $k$,
we also perform calculations for $\tau_0=0.8$ fm and $k=3$.  
The crucial point of the extrapolation from heavy ion to
light ion collisions is the use of the
$T$-dependent running $\alpha_s$ (\ref{eq:90}) with parameter $\kappa$ fitted
to heavy ion $R_{AA}$, that automatically
fixes $\alpha_s$ for light ion collisions.
Formula (\ref{eq:90}) for $\alpha_s$ with $c=0.8$ is motivated
by the lattice results \cite{Bazavov_al1}).
To understand the sensitivity of the results to a specific form of
the $Q$-dependence of $\alpha_s$ at $Q\lsim \kappa T$,
we also performed calculations
for paramtrization (\ref{eq:90}) with $c=0$, i.e., for a flat $\alpha_s$
at $Q<\kappa T$. 


\begin{figure} [!h] 
\begin{center}
\includegraphics[height=8.5cm]{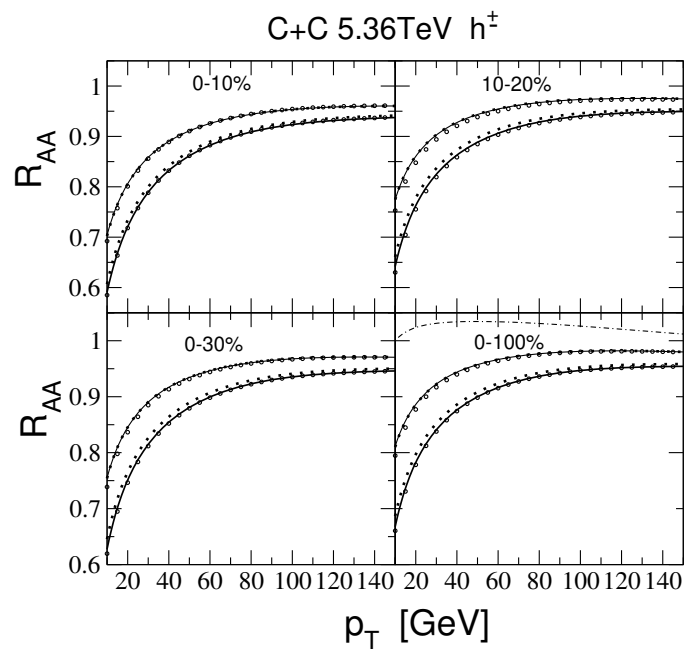}  
\end{center}
\caption[.]
{$R_{AA}$ of charged hadrons in $5.36$ TeV C+C collisions
  for the w/ mQGP and w/o mQGP scenarios (top to bottom).
  Solid and doted curves are for the parameter sets A and B (see table I),
  respectively.
  Circles show results for the parameter set C.
Long dash-dotted line shows $R_{AA}^{pdf}$.         }
\end{figure}
\begin{figure} 
\begin{center}
\includegraphics[height=8.5cm]{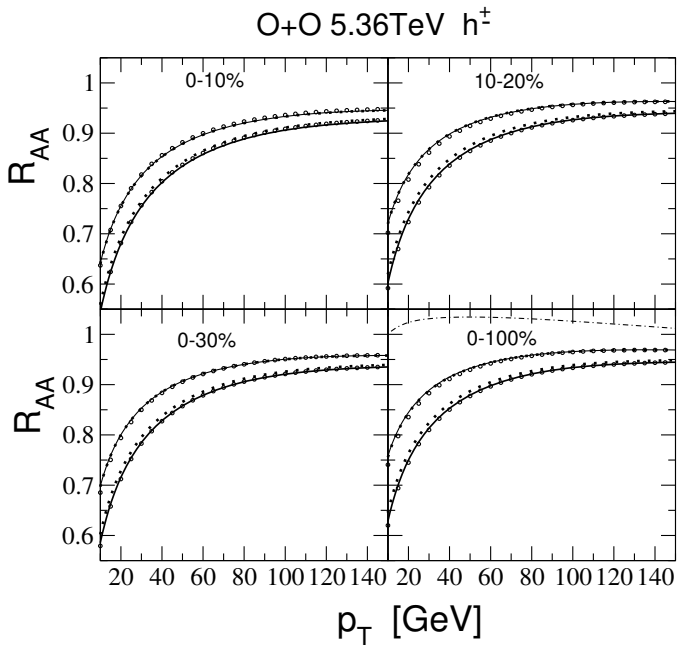}  
\end{center}
\caption[.]{Same as in Fig.1 for O+O collisions.
 }
\end{figure}
\begin{figure} 
\begin{center}
\includegraphics[height=8.5cm]{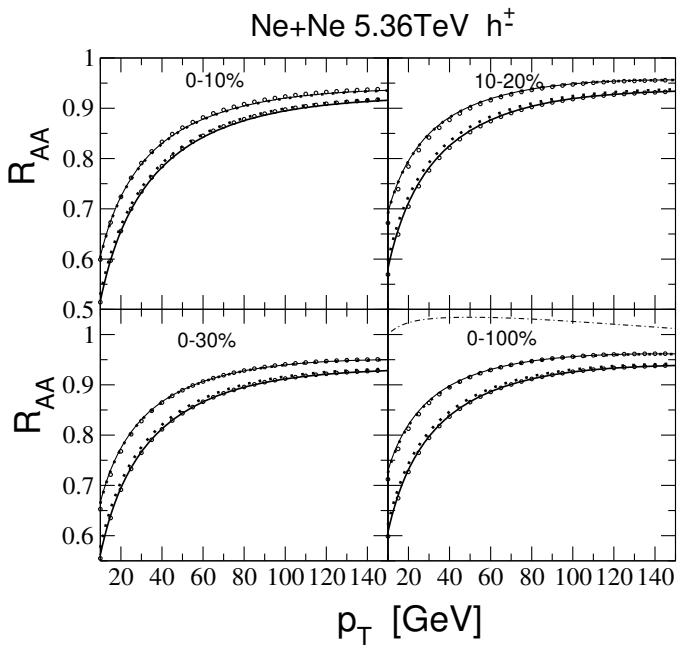}  
\end{center}
\caption[.]{Same as in Fig.1 for Ne+Ne collisions.
 }
\end{figure}

We perform calculations of $R_{AA}$ for four sets of $\tau_0$, $k$,
and $c$
listed in table I. There we also give the optimal
values of the parameter $\kappa$ fitted to data on $R_{AA}$
in 5.02 TeV Pb+Pb collisions in the w/ mQCD and w/o mQCD scenarios.
Comparing  the results for sets B, C, and D to that for A
allows to understand the magnitude of the theoretical
uncertainties of $R_{AA}$ related to the uncertainties:
in the thermalization time (B), in the geometry of the QGP fireball (C), and
in the $Q$-dependence
of $\alpha_s$ at $Q\lsim \kappa T$ (D).

\begin{table}
  \begin{tabular}{|c|c|c|c|c|}
    \hline
Set  & $\tau_0$ (fm)  & $\,\,k\,\,$ & $c$  & $\kappa$ \\
\hline
A &   0.5   &  2  & 0.8  & 2.35, 3.2\\
\hline
B &   0.8 &  2 &  0.8  & 2.25, 3.1\\
\hline
C &   0.5 &  3 &  0.8  & 2.7, 3.4\\
\hline
D &   0.5 &  2 &  0  & 2.55, 3.6\\
\hline
  \end{tabular}  
  \caption{
    Sets of the model parameters used for
    calculating $R_{AA}$ in 5.36 TeV C+C, O+O, and Ne+Ne collisions.
    The two numbers in the last column
     correspond to $\kappa$ fitted
    to the LHC data on $R_{AA}$ in 5.02 TeV Pb+PB collisions
    in the w/ mQGP and w/o mQGP scenarios, respectively.}
\label{param}
\end{table}

In Figs. 1--3 we show our results for the $p_T$ dependence of
$R_{AA}$ in 5.36 TeV C+C, O+O, and Ne+Ne collisions for
the parameter sets A (solid), B (dotted), and C (circles).
We show predictions for the w/ mQGP and w/o mQGP scenarios.
Also, we show in Figs. 1--3
the factor $R_{AA}^{pdf}$ calculated without the parton energy
loss in the QGP with the LO EPS09 \cite{EPS09} corrections to
the nucleon PDFs.
As one can see from Figs. 1--3,
the results for the sets B and C turn out to be very close to that
for our basic set A. We do not plot the results for the set D
because the curves for the sets A and D are practically indistinguishable.
Thus, we can conclude
that the extrapolation from the $R_{AA}$ in heavy ion collisions
to that in light ion collisions turns out to be quite stable to
possible  variations of the model parameters.

From Figs. 1--3 one sees that
the scenario with mini-QGP formation
in $pp$ collisions leads to larger $R_{AA}$, and
the difference
\bea
\Delta R_{AA}=R_{AA}(\text{\small w/ mQG}P)-R_{AA}(\text{\small w/o mQGP})
\nonumber\\
=R_{AA}^{st}(\text{\small w/ mQG}P)/R_{pp}-R_{AA}^{st}(\text{\small w/o mQGP})
\eea
becomes more pronounced with decreasing atomic number
(for minimum bias C+C and O+O collisions
we have $\Delta R_{AA}\sim 0.1-0.15$ at $p_T\sim 10-20$ GeV).
The mechanism of this effect is as follows.
For large nuclei the factor $R_{AA}^{st}$
is more sensitive to variation of $\alpha_s$ than in the case of light nuclei
 (due to larger QGP density and size for large nuclei).
This allows one to obtain for large nuclei $\Delta R_{AA}\approx 0$
with a small difference in $\alpha_s$ for the w/ mQGP and w/o mQGP scenarios.
It corresponds to
\bea
R_{AA}^{st}(\text{\small w/ mQGP})-R_{AA}^{st}(\text{\small w/o
  mQGP})\nonumber
\\
\approx 
R_{AA}^{st}(\text{\small w/o mQGP})
(R_{pp}-1)\,.
\eea
At the same time, for light nuclei (for which the sensitivity to $\alpha_s$
is weaker), such a small difference in
$\alpha_s$ leads to
$|R_{AA}^{st}(\text{\small w/ mQGP})-R_{AA}^{st}(\text{\small w/o mQGP})|$
that is considerably
smaller  than $R_{AA}^{st}(\text{\small w/o mQGP})(1-R_{pp})$.
As a result, for light nuclei one obtains
\beq
\Delta R_{AA}\sim R_{AA}^{st}(\text{\small w/o mQGP})(1-R_{pp})\,.
\eeq
This can lead to $\Delta R_{AA}\sim 0.1-0.15$ at
$p_T\sim 10$ GeV
(our calculations give $R_{pp}$ which grows from $\approx\!\! 0.78$ at
$p_T\approx 10$ GeV to $\approx\!\! 0.92$ at $p_T\approx 50$ GeV, we do not show the curve for $R_{pp}$
since it is
very close to that for $\sqrt{s}=5.02$ TeV shown in Fig.~3 of \cite{Z_hl}).

It is interesting to compare $|R_{AA}^{pdf}-1|$ and
$|R_{AA}-1|$ to see whether the uncertainties in $R_{AA}^{pdf}$
can be a serious obstacle for observation of jet quenching
effects, especially in the C+C collisions, for which $R_{AA}$ is the largest.
From Fig. 1 we observe that for minimum bias C+C collisions 
$|R_{AA}^{pdf}-1|/|R_{AA}-1|\lsim 0.2-0.3$  at $p_T\lsim 30$ GeV.
Thus, considering that the uncertainties in $|R_{AA}^{pdf}-1|$ do not
exceed $\sim 50$\%, we can conclude that for 5.36 TeV C+C collisions
the possible errors in
the nuclear PDFs should not have a strong impact on the jet
quenching interpretation of experimental data on $R_{AA}$ (at least in the
region $p_T\lsim 30$ GeV for minimum bias events).  

The final remark concerns the uncertainties in our predictions for
$R_{AA}$ for light nuclei that may come
  from errors in the Glauber model predictions for the
  midrapidity $dN_{ch}(AA)/d\eta$. Of course, these uncertainties
  are not related to the jet quenching model itself, and can be
  eliminated when data on $dN_{ch}(AA)/d\eta$ for light ions
  become available. Nevertheless, even without such data,
there are good reasons for the smallness of these uncertainties.
  Indeed, the fact that the Glauber model predictions of \cite{MCGL_Xe}
  for centrality dependence of $dN_{ch}(AA)/d\eta$ in 5.44 TeV Xe+Xe
  collisions for $dN_{ch}(AA)/d\eta\sim 50-200$ (at centrality $\sim 40-60$\%),
  which is of the order of the average multiplicity in jet events
  for C, O, and Ne nuclei, agree very well with
  the ALICE data \cite{ALICE_Nch_Xe}
  (the same is the case for ALICE data for 5.02 TeV Pb+Pb
  collisions \cite{ALICE_Pb5}),
  says that the errors
  of the Glauber model predictions for $dN_{ch}(AA)/d\eta$ for light nuclei
  cannot exceed a few percent. Such errors can lead to similar errors
  in the magnitude of $(1-R_{AA})$, i.e., the potential errors
  in $R_{AA}$ are $\lsim 0.01-0.02$ (at $p_T\lsim 50$ GeV),
  that is much  smaller
  than $\Delta R_{AA}$ in our predictions.


{\bf IV. Summary.}
We have calculated the nuclear modification
factor $R_{AA}$ in 5.36 TeV C+C, O+O, and Ne+Ne collisions
in scenarios with and without mini-QGP formation in $pp$ collisions.
We use the jet quenching scheme of \cite{Z_hl} based on the LCPI approach to
induced gluon emission, treating the collisional energy loss as a perturbation.
The radiative and collisional energy losses are calculated for
a temperature dependent running $\alpha_s(Q,T)$
with a plateau around $Q\sim \kappa T$
(motivated by the lattice results \cite{Bazavov_al1}).
We use values of $\kappa$ fitted to
the LHC data on $R_{AA}$
in 5.02 TeV Pb+Pb  collisions. This allows to make predictions
for $R_{AA}$ in light ion collisions without free parameters.
For the scenario with mini-QGP formation in $pp$
collisions we obtain $R_{AA}$ that is
noticeably larger than that in the scenario without mini-QGP formation
in $pp$ collisions. 
The difference in $R_{AA}$ between the scenarios
with and without mini-QGP formation in $pp$ collisions becomes more pronounced
with decreasing atomic number.
For minimum bias events in C+C and O+O collisions the
difference
between $R_{AA}$ in the scenarios with and without mini-QGP formation in
$pp$ collisions
is $\sim 0.1-0.15$ at $p_T\sim 10-20$ GeV.
This says that accurate experimental data on $R_{AA}$ in light ion collisions
could potentially give information on the presence of jet quenching
in $pp$ collisions.\\

\begin{acknowledgments}
  I am grateful to A.~Baty
  for useful communication on measurements
of $R_{AA}$ in 5.36 TeV light ion collision at the LHC
  by the CMS Collaboration. 	
\end{acknowledgments}

\end{document}